\documentclass[10pt,conference,a4paper]{IEEEtran}

\usepackage[utf8]{inputenc}

\usepackage{times}

\usepackage{graphicx}
\usepackage{subfigure}
\DeclareGraphicsExtensions{.png,.eps,.ps,.pdf}

\usepackage{url}

\usepackage{amsmath}
\usepackage[linesnumbered,lined,boxed,commentsnumbered]{algorithm2e}
\usepackage{listings}
\usepackage{graphicx}
\usepackage{siunitx}
\usepackage{dirtytalk}
\usepackage{verbatim}
\usepackage{listings}
\usepackage{color}
\usepackage{xcolor}
\usepackage{caption}
\usepackage{subcaption}

\definecolor{dkgreen}{rgb}{0,0.6,0}
\definecolor{gray}{rgb}{0.5,0.5,0.5}
\definecolor{mauve}{rgb}{0.58,0,0.82}

\RestyleAlgo{ruled}
\SetKwComment{Comment}{/* }{ */}
\SetKw{KwBy}{by}

\begin{document}
	\title{CTF as a Service: A Reproducible and Scalable Infrastructure for Cybersecurity Training}
	
	\author{\IEEEauthorblockN{Carlos Jimeno Miguel}
		\IEEEauthorblockA{Universidad Pública de Navarra\\
			Pamplona, Spain\\
			carlos.jimeno@unavarra.es}
		\and
		\IEEEauthorblockN{Mikel Izal Azcárate}
		\IEEEauthorblockA{Universidad Pública de Navarra\\
			Pamplona, Spain\\
			mikel.izal@unavarra.es}}

\maketitle

\begin{abstract}
	Capture The Flag (CTF) competitions have established themselves as a highly effective pedagogical tool for cybersecurity training, offering students hands-on experience in realistic attack and defence scenarios. However, organising and hosting these events requires considerable infrastructure effort that frequently limits their adoption in academic settings. This paper presents the design, iterative development, and evaluation of a CTF as a Service (CaaS) platform built on \textit{Proxmox} virtualisation, leveraging Infrastructure as Code (IaC) tools such as \textit{Terraform} and \textit{Ansible}, container orchestration via \textit{Docker Swarm}, and load balancing with \textit{HAProxy}. The system supports both a development-oriented workflow, in which challenges are automatically deployed from a Git repository through a CI/CD pipeline, and a deployment-oriented workflow for ad-hoc infrastructure provisioning. The design decisions adopted, the problems encountered during development, and the solutions implemented to achieve session persistence, external routing, and challenge replicability are described. The platform is designed to evolve towards a CTF hosting service with commercial potential, and future work directions in automatic scaling, monitoring integration, and frontend standardisation are outlined.
\end{abstract}


{\bf Contribution type:}  {Education and educational innovation}

\section{Introduction}

The growing global demand for cybersecurity professionals has intensified interest in innovative pedagogical approaches that equip students with the practical competencies required by industry. Traditional classroom instruction, while valuable for the acquisition of theoretical knowledge, is frequently insufficient to develop the offensive and defensive skills that characterise the profile of a security expert. In response to this need, academic institutions and professional organisations have increasingly adopted gamified competition formats as a complement to formal curricula.

Capture The Flag (CTF) competitions represent one of the most effective formats for cultivating cybersecurity competencies. A CTF is a cybersecurity competition in which participants must solve a series of technical challenges to obtain secret text strings called \textit{flags}, whose submission to the competition platform awards points to the team or individual who finds them. CTFs are typically organised in three main formats: the \textit{jeopardy} format, in which challenges are presented independently, classified by category and difficulty; the \textit{attack-defence} format, in which teams must simultaneously protect their own systems and compromise those of their adversaries; and the \textit{King of the Hill} (KotH) format, in which participants compete to gain control of a shared machine or system and hold it for as long as possible against all other competitors, combining offensive and defensive skills in real time. By presenting participants with realistic, self-contained challenges in areas such as reverse engineering, web exploitation, cryptography, forensic analysis, and binary exploitation, CTFs bridge the gap between theoretical knowledge and practical skills. Their game-like structure encourages participation, stimulates self-directed learning, and promotes the development of problem-solving strategies directly applicable to real-world scenarios.

The Universidad Pública de Navarra (UPNA) maintains a firm commitment to advancing cybersecurity education as part of a broader institutional strategy aimed at fostering technological innovation and supporting the development of projects with commercial transfer potential. With dedicated funding channelled towards educational initiatives and applied research, the university has actively sought to develop infrastructure capable of supporting recurring CTF events for its student community. This effort reflects the recognition that cybersecurity training is not only an academic priority, but also a driver of regional technological development and a pathway to skilled employment.

Organising and hosting a CTF event is, however, a technically demanding undertaking. Setting up isolated, reproducible environments for each challenge, ensuring equitable access for all participants, managing scaling under concurrent load, and maintaining reliable connectivity are operational challenges of considerable magnitude. In academic settings with limited technical staff and constrained hardware resources, the operational burden associated with CTF hosting can become a barrier that limits both the frequency and the quality of such events.

This paper addresses this challenge by presenting the design and iterative development of a CTF as a Service platform built on the UPNA's \textit{Proxmox} virtualisation infrastructure. By integrating IaC tools, container orchestration, and a CI/CD pipeline for challenge deployment, the system aims to reduce the operational burden of CTF hosting to a manageable level, enabling instructors and student organisations to run competitions with minimal manual intervention. The platform has been designed with extensibility as a core premise, with the long-term goal of evolving into a production service capable of supporting both internal university events and, potentially, external clients.

The remainder of this paper is organised as follows. Section~\ref{sec:soa} reviews related work on CTF platforms and cyber range systems. Section~\ref{sec:dev} describes the development process in detail, including the design decisions adopted and the technologies used to address them. Section~\ref{sec:results} presents the results obtained. Section~\ref{sec:future} outlines future work, and Section~\ref{sec:conclusion} closes the paper with conclusions.

\section{State of the Art}
\label{sec:soa}

\subsection{CTF Platforms and Competition Infrastructure}

The problem of hosting cybersecurity competitions has been addressed from multiple perspectives in the scientific literature. Wi \textit{et al.} \cite{wi2018gitctf} proposed \textit{Git-based CTF}, a lightweight approach to organising attack-and-defence classroom competitions that minimises operational costs for instructors. Their key contribution was identifying that existing version control infrastructure (Git repositories) could serve as the backbone for challenge distribution and solution submission, eliminating the need for dedicated competition servers. Although the proposal proved effective in the academic context of the Korea Advanced Institute of Science \& Technology (KAIST), it is inherently limited to small-scale scenarios in which challenge environments do not require dedicated containerised services or per-user isolation.

A broader perspective on the current state of CTF tooling is offered by Taylor and Arias \cite{taylor2024ctfstate}, who systematically evaluated the software ecosystem surrounding CTF competitions, distinguishing between game engines and challenge components. Their analysis identified a significant gap between static challenge platforms and dynamic infrastructure capable of supporting per-instance challenge environments. They conclude that most open-source game engines lack support for dynamic provisioning of isolated challenge instances, a limitation that the present work directly addresses.

Among open-source CTF management platforms, \textit{CTFd} \cite{ctfd} has become the de facto standard for competition logistics: team registration, scoreboard management, solution submission, and hint systems. Originally developed for the Cyber Security Awareness Worldwide (CSAW) competition at New York University, \textit{CTFd} prioritises ease of deployment and extensibility through a plugin architecture. However, the scope of \textit{CTFd} is limited to the competition management layer; it does not address the provisioning or lifecycle management of challenge infrastructure. Consequently, organisations using \textit{CTFd} must independently solve the problem of deploying and scaling the underlying environments.

\textit{Facebook's FBCTF} \cite{fbctf} represented a notable alternative due to its visually rich interface and world-map-style scoreboard. \textit{FBCTF} offered superior aesthetics compared to \textit{CTFd}, but proved significantly harder to install and maintain. As comparative evaluations note \cite{karagiannis2020}, \textit{FBCTF} is better suited for large-scale public competitions where visual impact is a priority, but its operational complexity makes it poorly suited to academic environments with limited DevOps resources. Furthermore, the project has been discontinued and no longer receives active maintenance.

A comprehensive study by Kucek and Leitner \cite{kucek2020survey} examined eight open-source CTF platforms and concluded that, while basic functionality was generally consistent across them, game configuration options vary substantially. The study highlights that no existing platform offered integrated support for dynamic challenge environments, per-user container isolation, or automatic scaling, confirming that the infrastructure layer remains a persistent gap in the CTF ecosystem.

\subsection{Cyber Ranges}

At the more sophisticated end of the spectrum, cyber range platforms such as \textit{KYPO} \cite{vykopal2017kypo} provide fully virtualised environments capable of simulating complex network topologies for training activities. Developed at Masaryk University since 2013, \textit{KYPO CRP} is built on \textit{OpenStack} and offers sandbox management, network emulation, and integrated monitoring capabilities. The platform has been used in training exercises with hundreds of participants and supports both educational and research use cases.

In the commercial and semi-professional domain, platforms such as \textit{Hack The Box} \cite{htb} and \textit{TryHackMe} \cite{thm} can equally be understood as cyber ranges oriented towards practical cybersecurity training: they offer online-accessible virtualised environments with machines and challenges that users attack in a controlled manner. Although their focus is individual training and their model is primarily subscription-based, they illustrate how the concept of a cyber range can be realised at different scales and for different purposes. The present work adopts this same broad understanding of the term, considering that a CaaS CTF hosting platform constitutes, in essence, a lightweight cyber range specialised in the competition format.

Nevertheless, platforms of the \textit{KYPO} class are architecturally complex and require considerable hardware resources, as well as specialised technical expertise for their deployment and operation. Their primary use case is large-scale cyber defence exercises simulating enterprise network environments, a considerably broader scope than what a \textit{jeopardy}-style CTF competition requires. The overhead associated with provisioning full virtual machines per challenge instance, as opposed to using lightweight containers, makes these platforms less suitable for the rapid, frequent, and cost-effective CTF hosting that the present work pursues.

\subsection{Summary and Positioning}

The existing landscape reveals a clear divide between lightweight competition management tools, which handle scoring and team logistics but ignore infrastructure, and full cyber range platforms, which offer rich virtualised environments but carry significant operational overhead. A space exists between these extremes for a platform that combines automated, reproducible challenge infrastructure with practical usability in resource-constrained academic settings. The present work occupies precisely that space, targeting an on-premises \textit{Proxmox} deployment with container-based challenges, automated CI/CD deployment, and a focus on operational simplicity.

\section{Development}
\label{sec:dev}
Before describing the architectural decisions and their implementation, it is useful to establish \textbf{what a CTF challenge is} from a technical standpoint, as this characterisation underpins many of the design decisions adopted.

A CTF challenge is a self-contained and reproducible unit composed of the source code of the vulnerable service, the resources needed to build it, and the configuration files that describe how to run it. In the context of this platform, challenges come from \textbf{trusted sources}: they are developed and maintained by our own collaborators and developers, which guarantees their quality, security, and pedagogical suitability. For participation in events, \textbf{guest users} are employed, whose access is confined to the competition environment without privileges over the underlying infrastructure. From that source code, a \textit{Docker} image is generated that encapsulates the complete challenge environment: the base operating system, dependencies, the running service, and the \textit{flag} that the participant must find. This image can be instantiated as a container on any system with \textit{Docker}, producing an identical execution environment regardless of the underlying hardware or operating system. Since each challenge is a self-sufficient unit that assumes nothing about the environment in which it runs, it is possible to deploy, replicate, stop, and replace challenge instances programmatically and without side effects on the rest of the infrastructure.

The platform is specifically designed for \textbf{\textit{jeopardy}-format CTFs}, which is the modality adopted by UPNA for its cybersecurity events. In this format, challenges are presented independently, classified by category (reverse engineering, web exploitation, cryptography, forensic analysis, binary exploitation, etc.) and difficulty level. Each challenge is autonomous and has an associated unique \textit{flag} whose correct submission awards points. Unlike the attack-defence format, participants do not interact with each other through the infrastructure, which considerably simplifies isolation and orchestration requirements.

The architecture described below aims to leverage the challenge format outlined above to provide a CTF competition hosting platform that is reproducible, scalable, and operationally straightforward, enabling the Universidad Pública de Navarra to organise cybersecurity events on a recurring basis without incurring a disproportionate administrative burden. To this end, the architecture integrates Infrastructure as Code tools that automate the provisioning and configuration of challenge environments, eliminating dependency on error-prone manual procedures. Through container orchestration, load balancing, and session persistence, the system ensures that each participant has access to an isolated, persistent, and always-available challenge environment, regardless of the load the platform is under at any given moment. Furthermore, the incorporation of a continuous integration and deployment pipeline allows challenge authors to publish and update their challenges continuously without needing to intervene on the underlying infrastructure, decoupling development from deployment.

\subsection{Architecture Design}

The starting point of the design was to establish a set of functional requirements that the platform must satisfy to be useful in a real competition context. The first and most fundamental is the \textbf{continuous availability of challenges}: each challenge must remain accessible throughout the entire duration of the event without requiring manual intervention, which implies that challenge environments must be resilient against individual container failures. It was therefore decided that each challenge would be deployed with multiple simultaneously running instances, so that the failure of one replica does not interrupt service for participants.

This decision immediately introduced the need for a \textbf{load balancer} to distribute incoming traffic among the available replicas of each challenge. Without an intermediary element to manage the distribution of connections, participants would need to know the address of a specific replica, which would negate the benefit of replication and complicate access management.

However, the replication of challenge instances is in direct tension with another key requirement: \textbf{preservation of participant progress}. Many CTF challenges, especially those in categories such as binary exploitation or system intrusion, maintain server-side state: a running process, a modified file, an active shell session. If the load balancer routes successive connections from the same participant to different replicas, the progress accumulated on the original replica is lost. Maintaining session affinity at the application level would be intrusive, as it would require modifying the code of each challenge to synchronise state between replicas, which is not feasible given that challenges are developed independently by different authors.

Another central design decision was to \textbf{confine each challenge to its own private network}, isolating it from other challenges and from general infrastructure traffic. This isolation is a basic security requirement: it prevents a participant who compromises one challenge environment from moving laterally to other challenges or to the management infrastructure. As a direct consequence, external traffic cannot reach challenge containers directly, as these reside in internal networks not routable from the Internet, requiring a \textbf{controlled entry point} to translate external connections towards the internal private networks.

\begin{figure}[htbp]
	\centering
	\includegraphics[width=\columnwidth]{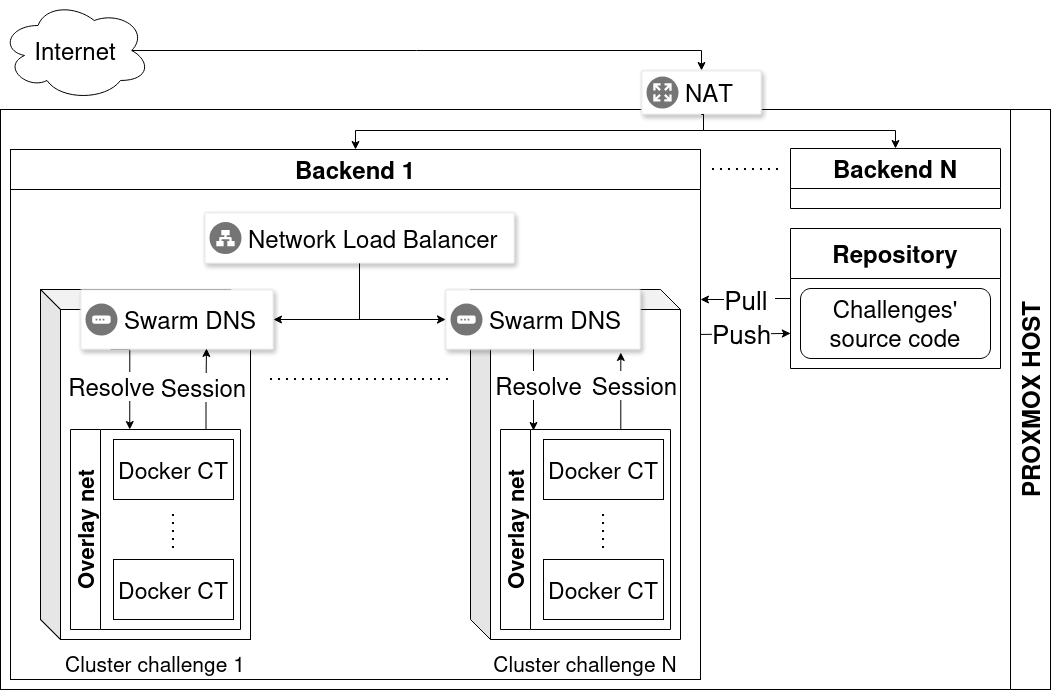}
	\caption{High-level architecture of the CaaS platform. Internet traffic enters through a NAT frontend and is distributed among backend nodes. Each backend hosts multiple container clusters (one per challenge), with a load balancer providing traffic distribution and session persistence.}
	\label{fig:arch1}
\end{figure}

Finally, the decision was taken to \textbf{separate infrastructure provisioning from service configuration}. Defining all infrastructure declaratively guarantees that environments are reproducible: the same set of configuration files must be able to generate an identical infrastructure at any time, eliminating dependence on informally documented manual procedures. Figure~\ref{fig:arch1} illustrates the architecture resulting from these decisions.

\subsection{Architecture Implementation}

\textbf{\textit{Proxmox} as the base hypervisor.} All infrastructure is deployed on \textit{Proxmox VE} \cite{proxmox}, an open-source hypervisor that allows managing both virtual machines and Linux Containers (LXC) from a single platform. This project uses LXC containers exclusively as the host-level isolation unit, as they offer faster startup and lower resource consumption than full virtual machines while maintaining a sufficient level of isolation for the different infrastructure roles (NAT frontend and backend nodes).

\textbf{\textit{Terraform} for declarative provisioning.} The creation of LXC containers in \textit{Proxmox} is managed via \textit{Terraform} \cite{terraform} through the community-maintained \textit{Telmate} provider. A modular structure was defined with independent modules for backend nodes (\texttt{backend\_lxc}), the NAT frontend node (\texttt{frontend\_lxc}), and resource pool management (\texttt{pool}). The \textit{Terraform} state is stored in a version control system, ensuring that the infrastructure is fully reproducible from the declarative specification and that all changes are recorded. \textit{Terraform} thus covers the requirement for separation between provisioning and infrastructure, but deliberately does not manage the configuration of the software running inside the containers, a responsibility that falls to \textit{Ansible}.

\textbf{\textit{Ansible} for node configuration.} Once the LXC containers have been provisioned, \textit{Ansible} \cite{ansible} handles their configuration through roles and playbooks. For backend nodes, tasks are defined for Docker installation, service user creation, and overlay network configuration. For the NAT frontend node, routing rules are managed. This separation of responsibilities between \textit{Terraform} and \textit{Ansible} keeps each tool within its natural domain and avoids the fragile patterns that arise when \textit{Terraform} attempts to take on configuration tasks through shell provisioners.

\textbf{\textit{Iptables} for external routing.} The NAT frontend node is configured with \textit{iptables} DNAT rules that translate incoming connections from the public network towards the corresponding backends. An \textit{Ansible} playbook (\texttt{iptables-conf.yml}) generates and applies these rules automatically from the infrastructure inventory, and persists them to survive system reboots. This mechanism resolves the controlled entry point problem identified in the design: external traffic arrives at the frontend, which forwards it to the appropriate backend without challenge containers needing to be exposed directly to the Internet.

\textbf{\textit{Docker Swarm} for container orchestration.} Within each backend node, challenges are deployed as \textit{Docker Swarm} \cite{dockerswarm} services rather than individual containers. Each service manages a set of replicas of the same challenge, directly addressing the continuous availability requirement: if a replica fails, \textit{Swarm} replaces it automatically. Services are configured with DNS Round Robin (DNSRR) endpoint mode, so that \textit{Swarm}'s internal DNS resolves the service name by cyclically returning the IP addresses of its active replicas. Each challenge resides in its own \textit{overlay} network managed by \textit{Swarm}, providing the inter-challenge isolation required by the design.

\textbf{\textit{HAProxy} for load balancing and session persistence.} \textit{HAProxy} \cite{haproxy} is deployed as a container on each backend, acting as a reverse proxy between the NAT frontend and the \textit{Swarm} services. For each challenge, \textit{HAProxy} queries the \textit{Swarm} DNS to obtain the addresses of all active replicas and distributes incoming connections among them using Round Robin. Session persistence, identified in the design as irreconcilable with application-level replication, is resolved here through \textit{HAProxy}'s \textit{stick-tables}: an in-memory table that associates each source IP address with the replica assigned at its first connection. As long as that replica remains active, all subsequent connections from the same IP are directed to it, preserving server-side state without any modification to the challenge code. The \textit{HAProxy} configuration is automatically generated via an \textit{Ansible} template (\texttt{haproxy.tpl}) populated with the metadata of the challenges deployed at any given time.

\begin{figure}[htbp]
	\centering
	\includegraphics[width=\columnwidth]{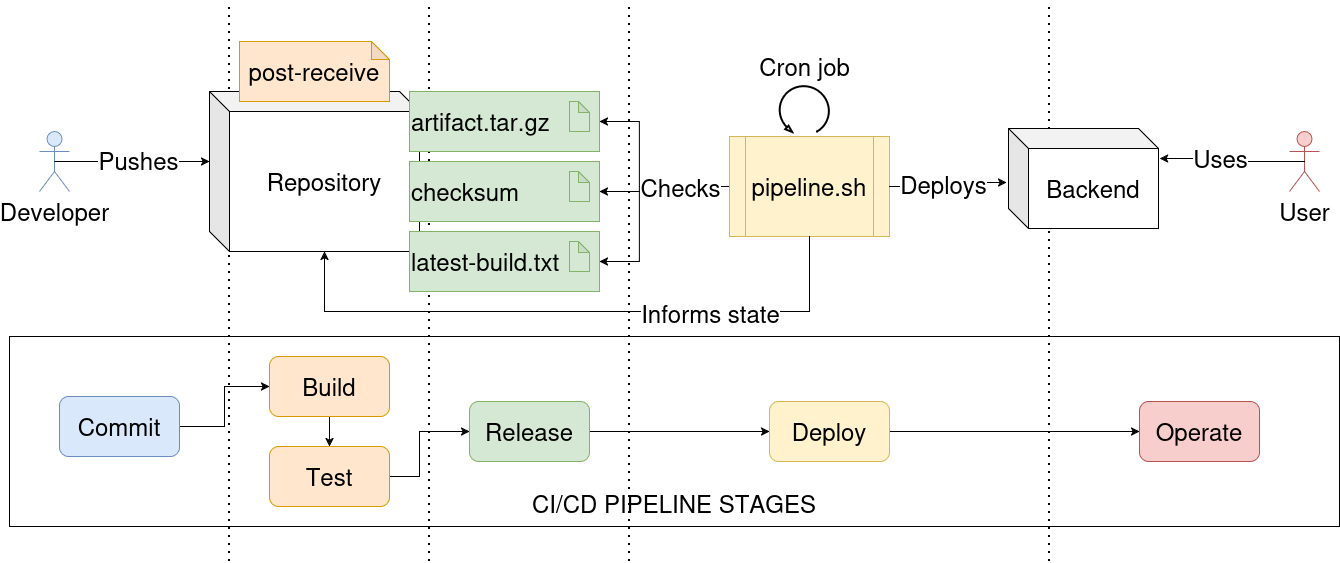}
	\caption{CI/CD pipeline flow.}
	\label{fig:cicd2}
\end{figure}

\textbf{CI/CD Pipeline and operational modes.} To automate the challenge lifecycle, a Git-based CI/CD pipeline was implemented, illustrated in Figure~\ref{fig:cicd2}.

A \texttt{post-receive} hook in the repository builds a compressed artifact with each push made by the developer and stores it in the \texttt{artifacts} branch. A \texttt{pipeline.sh} script, executed periodically as a cron job on each backend, checks for new artifacts, compares them with deployed versions, and updates the corresponding \textit{Swarm} services, reporting the result to the repository via a status file (\texttt{latest-build.txt}). This pipeline supports two operational modes: a \textbf{development-oriented mode} (\textit{rolling updates}), in which challenge authors publish changes that are automatically propagated to the backends, and a \textbf{deployment-oriented mode}, in which an administrator provisions an ad-hoc infrastructure from scratch by selecting a specific challenge catalogue.

\section{Results}
\label{sec:results}

The developed platform successfully achieves a series of functional objectives that address the operational challenges identified during earlier development phases.

\textbf{Reproducible infrastructure provisioning:} The combination of \textit{Terraform} and \textit{Ansible} allows a complete backend environment to be created from scratch in a repeatable manner. Starting from a freshly configured \textit{Proxmox} host, the entire infrastructure --- including the NAT frontend, the backend LXC nodes, Docker Swarm cluster initialisation, and HAProxy deployment --- can be brought up by executing a small number of defined scripts.

\textbf{Automated challenge deployment:} The Git hook-based CI/CD pipeline and the \texttt{pipeline.sh} script allow challenge authors to publish updates to the repository and have these automatically reflected in the active deployment without manual administrative intervention. The artifact-based approach decouples the build process from the deployment process, reducing the risk of partial or failed deployments.

\textbf{User session persistence:} The \textit{HAProxy} stick-table mechanism ensures that a participant connecting to a challenge is consistently directed to the same container replica throughout their session. This is essential for challenge types that maintain server-side state. Session affinity is maintained transparently without any modification to the challenge code.

\textbf{Challenge replicability and basic high availability:} \textit{Docker Swarm} services allow each challenge to be scaled to multiple replicas, providing basic load distribution and resilience against individual container failures. Although scaling currently requires manual administrator intervention (via \texttt{docker service scale}), the underlying mechanism is in place.

\textbf{External connectivity:} The \textit{iptables}-based routing chain and the \textit{HAProxy} reverse proxy provide a clean external connectivity path from the public Internet through the NAT frontend to the corresponding challenge containers, without challenges needing to expose ports directly on the host.

\textbf{Clear challenge format specification:} The platform defines a minimal contract for challenge packaging: a \textit{Docker} image built in the container registry, a \texttt{docker-compose.yml} specifying overlay network membership and DNSRR endpoint mode, and a pair of preparation and startup scripts adapted for \textit{Swarm} deployment. This contract is sufficiently minimal to be adopted without significant refactoring of existing challenges.

\section{Future Work}
\label{sec:future}

While the platform achieves its main functional objectives, several areas for improvement have been identified for subsequent development phases.

\textbf{Monitoring and observability:} The integration of Prometheus, started and subsequently paused in the early development phases, remains a priority. Metrics of interest include container availability, the number of connections per challenge, and resource utilisation per backend node. Beyond infrastructure metrics, developing \textit{Prometheus} probes capable of validating the correct functioning of deployed challenges (e.g., verifying that the service accepts connections and returns the expected initial response) would significantly improve operational visibility.

\textbf{Automatic scaling:} The current scaling mechanism requires manual administrator intervention to adjust the number of replicas. A demand-driven auto-scaling mechanism, potentially triggered by connection counts reported by \textit{HAProxy} or by \textit{Prometheus} resource utilisation metrics, would reduce the operational burden and improve participant experience during high-load periods. Integration with the \textit{Proxmox} API to dynamically provision additional backend LXC nodes when existing backends approach capacity would enable elastic scaling at the infrastructure level.

\textbf{Administration interface:} A web-based administration interface has been partially developed, although it is not yet in a definitive version. This interface would allow instructors without a technical background to manage challenge deployment, monitor participant activity, and configure competition parameters without direct command-line access. Integration with standard CTF management platforms such as \textit{CTFd} is also being explored, to provide participants with a familiar scoreboard and solution submission system.

\textbf{Challenge workflow standardisation and templates:} The current challenge format specification, while functional, could benefit from more rigorous standardisation. A challenge template repository with deployment scripts, a single configuration for continuous challenge integration, and documentation on how challenges should be designed would lower the barrier to entry for challenge authors and ensure consistency across the catalogue.

\textbf{\textit{HAProxy} high availability:} The current architecture deploys a single \textit{HAProxy} instance per backend, creating a potential single point of failure. A multi-instance \textit{HAProxy} deployment with shared state (via the \textit{HAProxy} \texttt{peers} configuration) and a floating virtual IP would eliminate this vulnerability.

\textbf{Improved deployment state tracking:} The current status notification mechanism stores a single deployment state per challenge without indicating which backend instance hosts it. Extending the \texttt{latest-build.txt} status file to include per-backend state would enable more granular monitoring and support multi-backend deployments in which different challenges are served from different backends.

\textbf{Integration with AI models:} the integration of challenges incorporating Artificial Intelligence models, both classical and generative, within the platform ecosystem is proposed. In particular, the development of orchestration mechanisms that enable the deployment of services based on language models (LLMs) as part of the challenges themselves or as auxiliary tools is envisioned. These systems could act as conversational assistants capable of providing dynamic hints, contextualized feedback, or progressive tutoring depending on the state of the challenge and the user's behavior. Additionally, the feasibility of integrating these services into the existing CI/CD pipeline will be studied, enabling the automatic updating of models or prompts associated with each challenge.

\section{Conclusions}
\label{sec:conclusion}

This paper has presented the design and iterative development of a CTF as a Service platform built on \textit{Proxmox} virtualisation using \textit{Terraform}, \textit{Ansible}, \textit{Docker Swarm}, and \textit{HAProxy}. The platform addresses the main operational barriers to hosting CTF competitions in academic settings: the complexity of infrastructure provisioning, the absence of automated deployment pipelines, and the difficulty of maintaining session persistence across replicated containers.

Over a series of development iterations, the system evolved from a manual deployment script to a cohesive, reproducible infrastructure stack with two differentiated operational modes supporting both development teams and deployment administrators. The resulting platform has demonstrated the feasibility of automated, Infrastructure as Code-based CTF hosting on own hardware, with a clear trajectory towards the commercial CTF as a Service offering that UPNA has as its horizon.

The platform constitutes a foundation upon which to build more advanced capabilities, including automatic scaling, integrated monitoring, and a polished administration interface, in future development cycles. The work demonstrates that open-source tools, when combined thoughtfully, can drastically reduce the operational burden of professional-grade CTF hosting.

\section*{Acknowledgements}

This work was supported by the Cybersecurity Chair established by the Universidad Pública de Navarra (UPNA) and its commitment to advancing cybersecurity education and fostering innovation in the field.

\nocite{*}
\bibliographystyle{IEEEtran}
\bibliography{bibliografia}

@inproceedings{wi2018gitctf,
	author    = {Wi, SeongIl and Choi, Jaeseung and Cha, Sang Kil},
	title     = {Git-based CTF: A Simple and Effective Approach to Organizing In-Course Attack-and-Defense Security Competition},
	booktitle = {2018 USENIX Workshop on Advances in Security Education (ASE 18)},
	address   = {Baltimore, MD, USA},
	month     = aug,
	year      = {2018},
	publisher = {USENIX Association},
	url       = {https://www.usenix.org/conference/ase18/presentation/wi}
}

@misc{taylor2024ctfstate,
	author       = {Taylor, G. M. and Arias, A.},
	title        = {CTF: State-of-the-Art and Building the Next Generation},
	howpublished = {Semantic Scholar},
	year         = {2024},
	url          = {https://api.semanticscholar.org/CorpusID:267660094}
}

@inproceedings{vykopal2017kypo,
	author    = {Vykopal, Jan and O{\v s}lej{\v s}ek, Radek and {\v C}eleda, Pavel and Vi{\v z}v{\'a}ry, Martin and Tov{\'a}rn{\'a}k, Daniel},
	title     = {KYPO Cyber Range: Design and Use Cases},
	booktitle = {Proceedings of the 12th International Conference on Software Technologies (ICSOFT 2017)},
	pages     = {310--321},
	year      = {2017}
}

@misc{ctfd,
	author       = {{CTFd Development Team}},
	title        = {CTFd: The Easiest Capture The Flag Platform},
	year         = {2024},
	url          = {https://ctfd.io/}
}

@misc{fbctf,
	author       = {{Facebook}},
	title        = {FBCTF: Facebook Capture the Flag},
	howpublished = {GitHub repository},
	url          = {https://github.com/facebook/fbctf}
}

@article{kucek2020survey,
	author  = {Kucek, Sebastian and Leitner, Michael},
	title   = {An Empirical Survey of Functions and Configurations of Open-Source Capture the Flag (CTF) Environments},
	journal = {Journal of Network and Computer Applications},
	volume  = {157},
	pages   = {102419},
	year    = {2020}
}

@incollection{karagiannis2020,
	author    = {Karagiannis, Stefanos and Maragkos-Belmpas, Emmanouil and Magkos, Efstathios},
	title     = {An Analysis and Evaluation of Open Source Capture the Flag Platforms as Cybersecurity e-Learning Tools},
	booktitle = {Information Security Education. Information Security in Action},
	publisher = {Springer International Publishing},
	address   = {Cham},
	pages     = {61--77},
	year      = {2020}
}

@misc{terraform,
	author = {{HashiCorp}},
	title  = {Terraform: Infrastructure as Code},
	url    = {https://www.terraform.io/}
}

@misc{ansible,
	author = {{Red Hat}},
	title  = {Ansible: Automation for Everyone},
	url    = {https://www.ansible.com/}
}

@misc{dockerswarm,
	author = {{Docker Inc.}},
	title  = {Docker Swarm Mode Overview},
	url    = {https://docs.docker.com/engine/swarm/}
}

@misc{haproxy,
	author = {{HAProxy Technologies}},
	title  = {HAProxy: The Reliable, High Performance TCP/HTTP Load Balancer},
	url    = {https://www.haproxy.org/}
}

@misc{proxmox,
	author = {{Proxmox Server Solutions GmbH}},
	title  = {Proxmox Virtual Environment},
	url    = {https://www.proxmox.com/}
}

@misc{htb, author = {{Hack The Box Ltd.}}, title = {{Hack The Box} --- Hacking Training For The Best}, year = {2025}, howpublished = {\url{https://www.hackthebox.com}}, note = {Consultado en 2025}}

@misc{thm, author = {{TryHackMe Ltd.}}, title = {{TryHackMe} --- Learn Cyber Security}, year = {2025}, howpublished = {\url{https://tryhackme.com}}, note = {Consultado en 2025}}
\end{document}